\documentclass[%
aps,
prb,
 amsmath,
 amssymb, 
reprint,%
numerical,
]{revtex4-1}
\usepackage{epstopdf} 
\usepackage{color}
\usepackage[english]{babel}
\usepackage{amsmath}
\usepackage{amsfonts}
\usepackage{graphicx}
\usepackage{array}

\begin{document}

\title{Electric Currents at Semiconductor Surfaces from the Perspective of Drift-Diffusion Equations}
\author{Jakub Lis}
\email{j.lis@uj.edu.pl}
\affiliation{Center for Nanometer-Scale Science and Advanced Materials (NANOSAM), Faculty of Physics, Astronomy and Applied Computer Science, Jagiellonian University, ul. St. Lojasiewicza 11, 30-348 Krakow, Poland}

\begin{abstract}
Surface sensitive electric current measurements are important experimental tools poorly corroborated by theoretical models. 
We show that the drift-diffusion equations offer a framework for a consistent description of such experiments. The current flow is calculated as a perturbation of an equilibrium solution depicting the space charge layer. 
We investigate the accumulation and inversion layers in great detail. Relying on numerical findings, we identify the proper length parameter, the relationship of which with the length of the space charge layer is not simple. If the length parameter is large enough, long-ranged modes dominate the Green's function of the current equation, leading to two-dimensional currents.
In addition, we demonstrate that the surface behavior of the currents is ruled by only a few parameters. This explains the fact that simplistic conductivity models have proven effective but makes reconstructions of conductance profiles from surface currents rather questionable.
\end{abstract}
\maketitle
\section{Introduction}
Electric current measurements at semiconductor surfaces at (sub-) micron distances have become a topic of interest\cite{popular1,popular2,popular3}. In such experiments, the electrostatic potential due to current injections from compact electrodes is locally probed at the surfaces. The progress in the construction of new multi-probe scanning tunneling microscopes~\cite{Wolkow,Marek} paves the way towards transport experiments at distances of hundreds (and perhaps tens) of nanometers. However, the understanding of the surface-sensitive current measurements is far from being satisfactory. In particular, there are several reports on the observation of two-dimensional current flows at semiconductor surfaces\cite{APL,JAP,2d1,2d2,2d3}. It is suggested that this is due to the formation of an inversion layer near the semiconductor surface. Alternatively, when two-dimensional currents are observed only at lower temperatures, decoupling from the bulk conductivity is argued, e.g., Ref.~\onlinecite{2d3}. A phenomenological approach widely explores the distinction between two- and three-dimensional currents~\cite{li} to describe the experimental results. However, the mechanism behind the generation of the two-dimensional currents and their relationship to the three-dimensional current flows are poorly understood.\\ 
In more analytical approaches, a two-step procedure is applied~\cite{popular1, popular2, popular3, PRB,nowe}. First, the variation of the carrier density related to the surface Fermi level pinning is calculated from the Poisson equation. Second, the current density is calculated using the (classical) drift current equation. Although this approach offers insight into the underlying physics, it imposes several difficulties. The most serious objection to the scheme results from the fact that the drift equation simply generalizes the Ohm's law. Consequently, it is not suitable for describing the p-n junctions present in inversion layers. Furthermore, the relation between the electrostatic potential behind band bending and the one behind the current flow remains elusive. An interpretation of their sum or difference has not been clarified. The last shortcoming of the scheme is that it does not take into account the recombination processes and it is unclear how they could be considered.\\
This motivated us to model the surface sensitive transport within a more general theory, which is the system of the drift-diffusion equations~\cite{jurgel}. These equations offer a unified description of the space charge layer and the transport experiments. In this paper, we restrict our model to two bands, i.e., electrons and holes. Extensions to other bands, e.g., surface bands, can be naturally worked out.\\
In addition, we pin down a mathematical mechanism responsible for the appearance of the two-dimensional currents within three-dimensional models. Our analysis reveals that only a few parameters are necessary to capture the surface manifestation of the non-trivial depth-dependent conductance; two parameters are sufficient in the case of inversion or accumulation layers. This explains why simple three-layer models of the conductivity profile can be successfully used to describe the experimental results\cite{voigt}. These findings are important as they shed light on the limitations of the reconstruction of the depth-dependent carrier profile from the surface-sensitive measurements~\cite{nowe}.\\
The paper is organized as follows. In the next section, we show how the drift-diffusion equations describe the space-charge layer and the current transport experiments. The model involves the Poisson equation for the electrostatic potential and current equations for every considered carrier type. Here, we formulate the current equations using a quasi-Fermi level Ansatz. In sect.~\ref{poissI}, we comment on the linearized Poisson equation and the reduction of the drift-diffusion equations to the drift equation. 
Next, in sect.~\ref{presentation}, we outline the presentation scheme for the surface profiles of the quasi-Fermi level and the voltage drop. The scheme facilitates the classification of solutions in terms of two- and three-dimensional currents as it takes advantage of the relevant scaling properties. 
Furthermore, the current equation is analyzed. After a brief summary of the known results (sect.~\ref{genFram}), we identify the parameters crucial for the surface currents and pinpoint the mechanism behind the appearance of two-dimensional currents (sect.~\ref{reduction}). These findings also enable comments on the reconstruction problem. 
In sect.~\ref{caseStudy}, we study the inversion layer at the Si(111)-Ag surface as reported in Ref.~\onlinecite{2d2}. Finally, we sum up our developments and point out a few open issues. 
Several technicalities are discussed in the Appendices\\
The equation for quasi-Fermi levels was examined with mathematical rigor in the context of the so-called Calderon problem~\cite{Calderon}. This theory puts the electrical impedance tomography~\cite{EIT} on firm ground. Despite some overlap between our work and these developments, these two problems are both physically and mathematically autonomous. \\

\section{Drift-diffusion model} 
\subsection{Model}\label{model}
Based on the Boltzmann transport equation for semiconductors, many calculation schemes of electric transport phenomena are derived~\cite{cosOgolnego,jurgel}, among them the drift-diffusion equations. An intuitive derivation from scattering analysis can be found in~Ref.~\onlinecite{Assad}. Notably, this theory is the first-choice scheme for device simulations~\cite{hansch}. \\
The (low current) drift-diffusion model is a coupled system of equations for the electrostatic potential (multiplied by the elementary charge $e$) $V(\mathbf{x})$ and electron $n(\mathbf{x})$ and hole $p(\mathbf{x})$ densities. The equations for stationary current flow read as
\begin{eqnarray}
\Delta V(\mathbf{x})=\frac{e^{2}}{\varepsilon_{0}\varepsilon_{r}}\left( n(\mathbf{x})-p(\mathbf{x})-N(\mathbf{x})\right),\label{napiecie}\\
\nabla\left(k_{B}T\mu_{n} \nabla n-\mu_{n} n \nabla V\right)=R(V, n,p)+f_{n}(\mathbf{x}),\label{prad1}\\
\nabla\left(k_{B}T\mu_{p} \nabla p+\mu_{p} p \nabla V\right)=R(V, n,p)+f_{p}(\mathbf{x}),\label{prad2}
\end{eqnarray}
where $N(\mathbf{x})$ stands for the net density of charged ions, $R$ for the recombination rate, $f_{n}(\mathbf{x})$ and $f_{p}(\mathbf{x})$ for external current sources of electrons and holes, respectively, $\varepsilon_{0}$ denotes the vacuum permittivity, $\varepsilon_{r}$ -- the material dielectric constant, $T$ -- the temperature, and $k_{B}$ -- the Boltzmann constant. The diffusion coefficient is absent in the equations due to the Einstein--Smoluchowski relation between the mobility $\mu$ and diffusion coefficient $D$, $D=k_{B}T\mu/e$. If the subscripts for the electrons and holes are not explicitly given, the relation holds for holes and electrons separately. On the right hand sides of eq.~(\ref{prad1}) and ~(\ref{prad2}), other current sources, like currents generated by the electromagnetic waves, can be added. The electron $\mathbf{j}_{n}$ and hole $\mathbf{j}_{p}$ current densities are 
\begin{eqnarray}
\mathbf{j}_{n}=\mu_{n}\left(k_{B}T\nabla n- n \nabla V\right),\\
\mathbf{j}_{p}=-\mu_{p}\left(k_{B}T\nabla p+ p \nabla V\right).
\end{eqnarray}
In equilibrium, all currents vanish: $\mathbf{j}_{n} =\mathbf{j}_{p}=0$. For electrons, the condition can be rewritten in the form
\begin{equation}
\nabla\frac{V_{0}}{k_{B}T}=\nabla \ln \frac{n_{0}}{n_{b}},
\end{equation}
and hence 
\begin{equation}
n_{0}(\mathbf{x})=n_{b} \exp{\frac{V_{0}(\mathbf{x})}{k_{B}T}}.
\end{equation}
The constant electron density $n_{b}$ is introduced to have a dimensionless expression under the logarithmic function. An analogous calculation can be performed for holes. The above calculation makes it evident that the equations~(\ref{napiecie})--(\ref{prad2}) build on the Boltzmann statistics.
The space-charge layer is modeled by the appropriate boundary condition imposed on $V$~\cite{Monch,InvProb}. We denote the equilibrium potential and carrier densities with a subscript zero. These equilibrium quantities depend on the distance from the surface only and we assume that the carrier densities saturate in the bulk. Below, we will consider the convention that the surface is located at $z=0$ and the semi-axis $z>0$ corresponds to the crystal bulk. \\
The surface transport experiments are believed to slightly perturb the structure of the space charge layer without any substantial damage to it. As such, we are interested in the linearization of the above equations around the $V_{0}$, $n_{0}$, and $p_{0}$ solutions. To this end,   we write the electrostatic potential in the following form
\begin{equation}
V=V_{0}+v,
\end{equation}
and parametrize the carrier densities with the quasi-Fermi levels $\varphi$,
\begin{eqnarray}\label{quasiFermin}
n=n_{0}\exp\frac{v+\varphi_{n}}{k_{B}T},\\
p=p_{0}\exp\frac{-v-\varphi_{p}}{k_{B}T}. \label{quasiFermip}
\end{eqnarray}
 A quasi-Fermi level traces the deviations from the equilibrium occupancy of the related band. The linearized equations are obtained by plugging the above Ansatz into eq.~(\ref{napiecie})--(\ref{prad2}) and keeping the linear terms in $v$, $\varphi_{n}$, and $\varphi_{p}$ only. The final relations read as
\begin{eqnarray}
-\Delta v+\lambda\left( n_{0}+p_{0}\right)v=-\lambda \left(n_{0}\varphi_{n}+p_{0}\varphi_{p}\right),\label{poisson}\\
\nabla\left(\mu_{n}n_{0} \nabla \varphi_{n}\right)=\tilde{R}(V_{0},v,\varphi_{n},\varphi_{p})+f_{n},\label{lincur1}\\
-\nabla\left(\mu_{p} p_{0} \nabla \varphi_{p} \right)=\tilde{R}(V_{0},v,\varphi_{n},\varphi_{p})+f_{p}\label{lincur2},
\end{eqnarray}
where $\tilde{R}$ stands for the linearized recombination rate, and 
\begin{equation}
\lambda=\frac{1}{k_{B}T} \frac{ e^{2}}{\varepsilon_{0}\varepsilon_{r}}.
\end{equation} 
A similar approach can be found in Ref.~\onlinecite{InvProb}.
In the following, we neglect band mixing $\tilde{R}$ and this results in the decoupling of the three equations: the current equations can be calculated independently and then the Poisson equation can be solved. 
The boundary conditions imposed on any of the quasi-Fermi levels $\varphi$ ensure that there are no currents flowing through the surface: 
\begin{equation}\label{boundaryEq}
\left.\frac{\partial}{\partial z}\varphi(\mathbf{x}) \right|_{z=0}=0.
\end{equation}
We assume that the surface charge is not substantially changed by the transport experiments, and hence 
\begin{equation}
\left.\frac{\partial}{\partial z} v(\mathbf{x}) \right|_{z=0}=0.
\end{equation}
Upon the identification $\sigma_{n}=\mu_{n} n$ (and analogously for holes), the current equations~(\ref{lincur1})--(\ref{lincur2}) get the form of the (classical) drift current equation. Hence, previous results~\cite{nowe,PRB} can be straightforwardly adapted to the framework of the drift-diffusion equations. 
At this stage, one could incorporate surface bands into the theory by adding an additional band to eq. (\ref{napiecie})--(\ref{prad2}), but  we leave this issue to a future study. \\
The system of drift-diffusion equations has been widely used in electronic device simulations, but one should keep in mind that it is an approximation. The model does not report on thermal effects, and so it is not suitable for systems with substantial heating of the sample. Also, the equations are based on the quasi-classical description of the electronic bands and the Boltzmann statistics for charge carriers. Hence, they may not be able to capture quantum effects and poorly perform for systems with Fermi levels close to the band edges. Models consistent with the Fermi--Dirac statistics~\cite{jurgel} have rarely been considered in the literature. We believe that the qualitative analysis delivered below remains relevant even for the Fermi--Dirac statistical distribution of the charge carriers.

\subsection{Linearized Poisson equation \label{poissI}}
The linearized Poisson equation has the structure of the Schr{\"o}dinger operator with a source. The homogeneous part of the equation reads
\begin{equation}
\underbrace{\left[-\Delta +\lambda\left( n_{0}+p_{0}\right)\right]}_{\hat{P}}v=0.
\end{equation}
The potential is positive and reflects the screening efficiency of the free charge carriers. The inverse of $\sqrt{\lambda(n_{0}+p_{0})}$ is called the Debye--H{\"u}ckel length. It is instructive to consider perturbations around a constant carrier density $\varrho$. Then, the equation turns into the well-known screened Poisson equation. A Fourier transformation of the equation $\hat{P}v=-\lambda \varrho \varphi$ yields the relation
\begin{equation}
\tilde{v}(k)=-\frac{\lambda \varrho\tilde{\varphi}(k)}{k^{2}+\lambda\varrho},
\end{equation}
where the transformed functions are denoted with tildes.
If the source term $\tilde{\varphi}(k)$ has a considerable amplitude for small $k$ ($k^{2}\ll\lambda\varrho$) only, then the approximate solution reads as
$ \tilde{v}=-\tilde{\varphi}$, and hence $v=-\varphi$. Plugging this into the linearized current equation, we obtain
\begin{equation}\label{drift}
\nabla\left(\sigma\nabla v\right)=\pm f,
\end{equation}
where the sign of $f$ depends on the band under consideration. 
It is the drift equation widely used in modeling the surface current measurements. Here, it emerges as a long-wavelength approximation to the drift-diffusion equation. It is valid far from the current source, where the quasi-Fermi levels are determined by small modes. This can also be seen from the Green's function in real space, which reads
\begin{equation}
v(\mathbf{x})=-\lambda \varrho \int d^{3}y \ \frac{\exp\left(-\sqrt{ \lambda\varrho} |\mathbf{x}-\mathbf{y}|\right)} {|\mathbf{x}-\mathbf{y}|}\varphi(\mathbf{y}).
\end{equation}
This equation shows that the potential $v$ can be regarded as the smoothed-out quasi-Fermi level.
For the slowly varying charge densities, the abovementioned Green's function can be used approximately upon the substitution $\varrho\to\varrho(\mathbf{x})$ (adiabatic approximation). The characteristic length-scale dividing the short- and long-ranged variations is given by $1/\sqrt{\lambda \varrho}$. For silicon and germanium, we give the values of the Debye--H\"uckle length in Tab.~\ref{tab1}. For large carrier densities, the parameter is in the range of several nanometers, and the drift equation appears as a legitimate approximation. At low carrier densities or quick variation of $\sigma$, this is not true and one needs to deal with the equation system. \\
\begin{table}
\begin{tabular}{c||c|c||c |c}
 & \multicolumn{2}{c||}{Si} &\multicolumn{2}{c}{Ge} \\
 \hline
 $\varrho$ [cm$^{-3}$] &100 K & 300 K &100 K & 300 K \\
 \hline \hline
$10^{13}$ & 720 nm & 1244 nm & 847 nm & 1467 nm \\
$10^{16}$& 23 nm & 39 nm  & 27 nm & 46 nm \\
$10^{18}$& 2.3 nm & 3.9 nm & 2.7 nm & 4.6 nm \\
\end{tabular}
\caption{Debye--H{\"u}ckel length $(\lambda \varrho)^{-1/2}$ for silicon and germanium at various temperatures and carrier densities $ \varrho$.\label{tab1}}
\end{table} 
The intuition that the electrostatic potential smears the quasi-Fermi levels over a certain volume is well known. As such, one can expect that the notions of two- and three-dimensional currents and electrostatic potentials remain valid within the drift-diffusion equations. Indeed, if a current equation results in the quasi-Fermi level behaving at the surface as $\ln(r)$ or $r^{-1}$, then the electrostatic potential follows the same function with a different multiplicative factor ($r$ stands for the distance from the current source). Notably, while the quasi-Fermi level associated with the logarithmic function is $z$-independent, the electrostatic potential will vary with the distance from the surface. More details are given in Appendix~\ref{poissII}. 

\section{Presentation scheme\label{presentation}}
The distinction between two- and three-dimensional currents marks a qualitative difference. In the first case, $\varphi$ and $v$ at the surface vary as $\ln(r)$, in the latter as $r^{-1}$. For the sake of a qualitative discussion focused on this distinction, related scaling properties are helpful. The presentation scheme applied in recent experimental reports~\cite{APL,JAP} makes these relations manifest. In those experiments, the current source and drain were located at the surface at $(0,0)$ and $(D,0)$, and two additional electrodes measured the voltage drop between points $\tfrac{D}{2}(1+x,0)$ and $\tfrac{D}{2}(1-x,0)$ for some $0<x<1$. To illustrate this, we show the resistance $R$ (voltage drop divided by the current flowing through the system) as a function of $x$ for several $D$. Then~\cite{APL}, the resistance $R$ depends only on $x$ in the case of two-dimensional currents. Both $x$ and $D$ are necessary to determine the resistance in the three-dimensional case where the quantity $R\cdot D$ is independent of $D$. In both cases, the formulae for the resistance are analytic:
\begin{equation}\label{2d}
R_{2}=\frac{v_{2}(0)}{\pi\sigma_{2}}\ln\frac{1+x}{1-x}
\end{equation}
for two-dimensional currents, where $\sigma_{2}$ is a two-dimensional conductivity parameter, and 
\begin{equation}\label{3d}
R_{3}=\frac{v_{3}(0)}{ D\pi\sigma_{3}}\frac{x}{1-x^{2}}
\end{equation}
for three dimensional-currents; $\sigma_{3}$ stands for the three-dimensional conductivity. The numerical factors, $\sigma_{2}$ and $\sigma_{3}$ result from the current equation, while $v_{2}(0)$ and $v_{3}(0)$ from the Poisson equation, see Appendix~\ref{poissII}. Obviously, if $v_{2}=v_{3}=1$, one gets the formulae obtained from the drift equation.\\

\section{Current equation}
\subsection{General framework \label{genFram}}
Now, we consider the current equations (\ref{lincur1}) and~(\ref{lincur2}). As already mentioned, we neglect the recombination and, in consequence, the current equations for different bands get decoupled. We concentrate on the relation
\begin{equation}\label{start}
\nabla\left(\sigma\nabla\varphi\right)=f,
\end{equation}
where $\varphi$ is the quasi-Fermi level, $\sigma>0$ stands for the conductivity, and $f$ is a source function. The boundary condition~(\ref{boundaryEq}) prevents any current flow through the surface. A comparison of eq.~(\ref{start}) and eq.~(\ref{lincur1})--(\ref{lincur2}) makes it clear that the conductivity characterizes every single band. We aim for a description of the quasi-Fermi level close to the surface and the extraction of physical content. We begin with a brief r{\'e}sum{\'e} of the results published in Ref.~\onlinecite{PRB}. In that paper, it was pointed out that the theory may result in strange long-ranged behavior if $\tfrac{d}{dz}\sigma(0)\neq 0$. We clarify this issue in Appendix~\ref{app}, concluding that there is no physical difference between $\tfrac{d}{dz}\sigma(0) = 0$ and $\tfrac{d}{dz}\sigma(0) \neq 0$.

To arrive at the general solution of eq.~(\ref{start}), we look for solutions of the function $\xi$:
\begin{equation}\label{transform}
\xi=\sqrt{\sigma} \varphi,
\end{equation}
which recasts eq.~(\ref{start}) into a Schr{\"o}dinger-like form:
\begin{equation}\label{operatorL}
\underbrace{\left[-\Delta+\frac{\Delta \sqrt{\sigma}}{\sqrt{\sigma}}\right]}_{\hat{L}}\xi=\sigma^{-1/2}f.
\end{equation}
The boundary condition~(\ref{boundaryEq}) is
\begin{equation}\label{boundaryTransformed}
\left.\frac{d\xi(z)}{dz}\right|_{z=0}=\left.\frac{1}{2\sigma}\frac{d\sigma(z)}{dz}\right|_{z=0}.
\end{equation}
The useful formula for the Green's function $G$ is given in terms of solutions of the following one-dimensional equation 
\begin{equation}\label{1dim}
\left(-\frac{d^{2}}{dz^{2}}+U(z)\right)\psi(k;z)=k^{2}\psi(k;z),
\end{equation}
where the potential $U(z)$ is given by the formula
\begin{equation}\label{potential}
U(z)=\sigma^{-1/2}\frac{d^{2}\sqrt{\sigma}}{dz^{2}}.
\end{equation}
$\{\psi(k;z)\}_{k}$ are normalized (generalized) eigenfunctions of the linear operator $\hat{L}$. The general solution of eq.~(\ref{start}) is 
\begin{equation}
\varphi(\mathbf{x})=\int dx'dy'dz' G(\mathbf{x};\mathbf{x'})f(\mathbf{x'}),
\end{equation}
where 
\begin{equation}\label{green}
\begin{split}
G(\mathbf{x};\mathbf{x'})=\frac{1}{2 \pi \sqrt{\sigma(z)\sigma(z')}} \\ \times \int_{0}^{\infty} dk \ K_{0}(k\sqrt{(x-x')^{2}+(y-y')^{2}})\psi(k;z)\psi(k,z').
\end{split}
\end{equation}
$K_{0}$ stands for the modified Bessel function of the second kind of the zeroth order. 
As demonstrated in Appendix~\ref{opL}, operator $\hat{L}$ has a positive spectrum only; in the case of semi-space, it corresponds to the continuous spectrum.
We consider below the quasi-Fermi level due to a point source located at $(x'', y'' ,z''=0)$
\begin{displaymath}
f(\mathbf{x'})=I\delta(x'-x'')\delta(y'-y'')\delta(z'),
\end{displaymath}
where $I$ stands for the current supplied to the sample. This gives rise to the formula for the surface profile $\phi(r)=\varphi(r,z=0)$: 
\begin{equation}\label{integ}
\phi(r)=\frac{I}{2\pi\sigma(0)} \int_{0}^{\infty} dk \ K_{0}(kr)\psi^{2}(k;0),
\end{equation}
where $r$ is the two-dimensional radius $\sqrt{(x-x'')^{2}+(y-y'')^{2}}$. Below, we consider a one-source model, although it leads to sample electrical charging (no stationary solution). A complete model has to include both current sources and drains. Due to the linearity of the equations, this is a simple generalization of the one-source case.\\
This paper deals with the elementary features of the model. In particular, we are interested in what can be seen on surfaces at some distance from the source. We do not examine the structure of the very contact and the details of the injection. This approach is justified by the local character of the screened Poisson equation and the fact that two- and three-dimensional currents depend on the long-ranged modes. \\

\subsection{Mechanism behind the dimensional reduction\label{reduction}}
\begin{figure*}[t]
\includegraphics[scale=0.35]{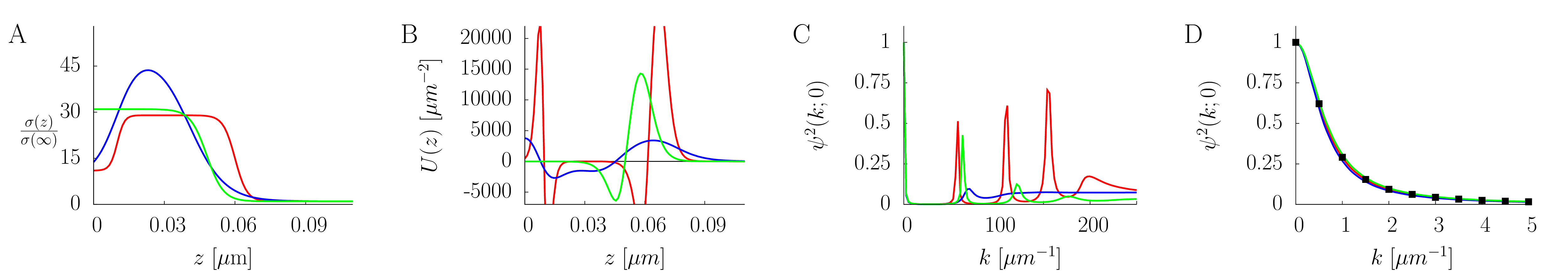}
\caption{Panel A shows the three conductivity profiles numerically considered: $ 14(1-\tanh \tfrac{z-0.06}{0.005})-9(1-\tanh \tfrac{z-0.01}{0.003})+1$---red, $25(1-\tanh \tfrac{z-0.0405}{0.015})-21(1-\tanh \tfrac{z-0.01}{0.01})+1$---blue, and $15(1-\tanh \tfrac{z-0.048}{0.007})+1$---green. The resulting potentials $U(z)$ are shown in panel B; the following functions $\psi^{2}(k;0)$ in panel C. Image D is a close-up view of $\psi^{2}(k;0)$ for small $k$ with several points corresponding to the Lorentzian, as defined in eq.~(\ref{lorentz}), with $\Gamma^{2}=0.41 \ \mu m^{-2}$, $p=0$, and $A=1$ shown for illustration. The functions $\psi^{2}(k;0)$ are rescaled so that $\psi^{2}(0;0)=1$ (panels C and D).\label{potent}}
\end{figure*}
\begin{figure*}[t]
\includegraphics[scale=0.45]{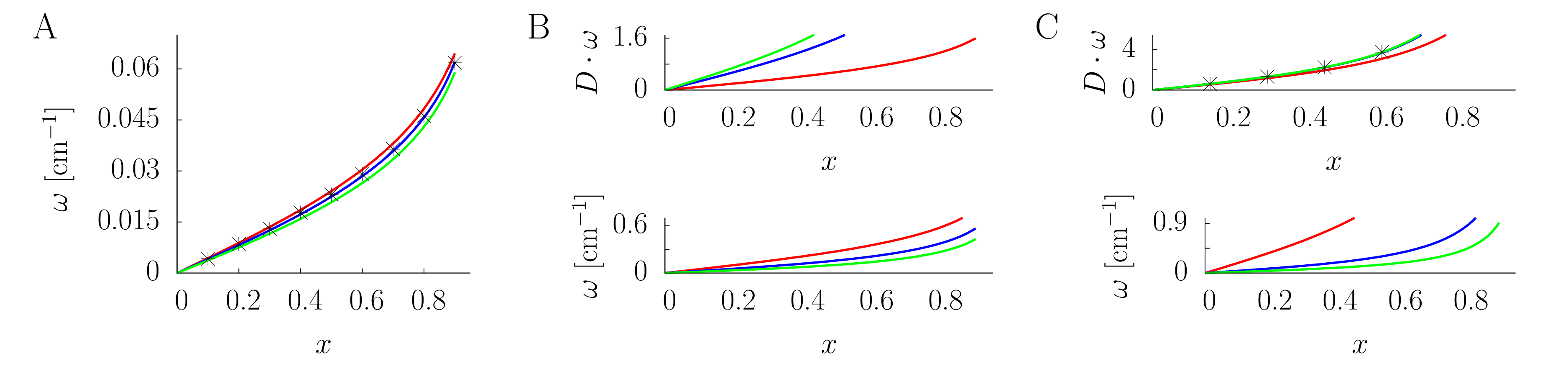}
\caption{Normalized resistance $\omega= \sigma(\infty) \phi/I$ for the geometry depicted in sect.~\ref{presentation} for various $\Gamma^{2}$: $\Gamma^{2}=0.0005 \ \mu m^{-2}$ in A, $\Gamma^{2}=0.1 \ \mu m^{-2}$ in B, and $\Gamma^{2}=20 \ \mu m^{-2}$ in C. Colors correspond to different values of $D$: $2\ \mu m$---green, $10\ \mu m$---blue, and $20\ \mu m$---red. The two-dimensional character resulting in the independence of $D$ is shown in panel A. The nearly three-dimensional case is shown in panel C as the quantity $DR\sigma(\infty)=D\omega$ appears to be $D$-independent. An exemplary transition behavior is shown in panel B, where neither graphs of $\omega(x)$ nor $D\omega(x)$ coincide. The points (stars) in panels A and C demonstrate the validity of eq.~(\ref{2d}) and eq.~(\ref{3d}), respectively with fitted numerical coefficients. $D\omega$ is a dimensionless quantity while $\omega$ has the dimension $length^{-1}$, we write $cm^{-1}$ due to the convention $[\sigma]=(\Omega cm)^{-1}$.\label{resist}}
\end{figure*}

The following analysis builds on eq.~(\ref{integ}). It clarifies that the conductivity profile impacts the surface current measurements through the function $\psi^{2}(k;0)$. As such, we numerically investigate the possible shapes of that function for different profiles $\sigma(z)$. In Fig.~\ref{potent}A, three close but clearly distinct conduction profiles~\cite{note} are shown, in Fig.~\ref{potent}B---resulting potentials $U(z)$ and in Figs.~\ref{potent}C and D--- functions $\psi^{2}(k;0)$. The most important fact about $\psi^{2}(k;0)$ is evident at the first glance, i.e., the functions have several pronounced peaks, in particular around $k=0$. In general, the peaks seem to originate from resonances; shape resonances naturally appear in the potentials of the type shown in Fig.~\ref{potent}C with regions, from which a classical particle cannot escape unless it is given sufficient energy. It can be inferred that the resonances mark effective transmission channels between the surface and the bulk through the potential $U$. 
On a semi-axis, where the Laplace operator is not always a positive operator~\cite{Bonneau}, the zero-resonance is allowed even for a positive and monotonically decreasing potential $U(z)$. We observe the zero-resonance if $\sigma$ approaches its bulk value from above, i.e., there is some enhancement of the conductivity close to the surface. In the opposite case, there is a zero-antiresonance and $\psi^{2}(k;0)$ has a local minimum for $k=0$. We aim to describe experiments at large distances, so we concentrate on the peak for $k=0$; however, analogous discussions can be conducted for any other peak.\\
As shown in Fig. \ref{potent}D, the $k=0$ peaks calculated for three different functions $\sigma$ are nearly indistinguishable and well approximated by a Lorentzian. Based on our experience, this is generic behavior: any well-defined peak of $\psi^{2}(k;0)$ is satisfactorily reproduced by the Lorentz function
\begin{equation}\label{lorentz}
g(\Gamma,A,p;k)=\frac{A\Gamma^{2}}{(k-p)^{2}+\Gamma^{2}},
\end{equation}
where $A$ stands for the peak amplitude, $\Gamma$ for its width, and $p$ for its position. Discrepancies are sometimes observed at the tails. \\
 Following formula~(\ref{integ}), $\phi(r)$ emerges upon integration. Function $K_{0}(x)$ diverges at the origin and quickly vanishes starting from $x\approx 3$. A thin peak in $\psi^{2}(k;0)$ can dominate the integral so that we can approximate 
\begin{equation}
K_{0}(kr)\sim -\ln{kr}=-\ln{\frac{k}{q}} - \ln{q r}, \label{Klog}
\end{equation}
where $q$ is an arbitrary parameter. This allows writing the quasi-Fermi level at the surface in the form
\begin{equation}\label{log}
\phi(r)\approx-\int_{0}^{\infty} dk \ \left(\ln{\frac{k}{q}} + \ln{q r}\right) \psi^{2}(k;0), 
\end{equation}
and hence
\begin{equation}
\phi(r)=const + \ln{(qr)} \int_{0}^{\infty} dk \ \psi^{2}(k;0).
\end{equation} 
The constant term can be absorbed into the logarithm modifying $q$. This parameter has no physical meaning as it corresponds to shifting the potential by a constant value, and when considering both current source and drain, one can get rid of this parameter. Note that there is no such parameter in eq.~(\ref{2d}), which corresponds to the physical measurements. So far we have taken into account the peaked structure of $\psi^{2}(k;0)$ with no assumption as to its functional form. If $\psi^{2}(k;0)$ is given by a Lorentzian, the integral can be performed, and the two-dimensional (sheet) conductivity can be expressed as
\begin{equation}\label{sigma2}
\sigma_{2}=\frac{\sigma(\infty)}{\pi^{2}\Gamma}.
\end{equation} 
The above formula takes advantage of the fact that $\psi^{2}(0;0)=\sigma(0)/\sigma(\infty)$~\cite{PRB}. 
If $r$ is too large ($0.1 \lesssim\Gamma^{2} r^{2}$ for a Lorentzian), the peak becomes broad and the approximation~(\ref{Klog}) is no longer instrumental. Then, three-dimensional conductivity prevails~\cite{PRB}. An Ansatz with a single peak for $\psi^{2}(k,0)$ allows the modeling of the two- and three-dimensional character of the surface currents, as demonstrated in Fig.~\ref{resist}. Notably, parameter $\Gamma^{-1}$ is the only characteristic length and its value is not simply related to the width of the space charge layer or any other length parameter to be identified in the system. \\
The logarithmic solution of eq.~(\ref{start}) appears for a plate as a zero-mode with no $z$-dependence. It dominates the solution for $r$ comparable to the plate thickness. We argue in Appendix~\ref{const} that the quasi-Fermi levels are approximately $z$-independent, even in the semi-space. It follows from the fact that the long-ranged modes closely resemble the zero-mode beneath the surface.\\
The above considerations show how the confined currents appear in the system and allow calculations of the two-dimensional conductivity from the function $\sigma(z)$. The physical significance of the structure of the theory goes beyond this statement. Assuming the Lorentz shape of the peak around $k=0$, we see that the surface conductivity far from the source is governed by two parameters ($\Gamma$, $\sigma(\infty)$) only. At smaller distances,  additional three parameters can appear: the position, width, and amplitude of the second peak. Typically, the second peak becomes important at distances well below 1~$\mu$m and the third one at distances where the theory is no longer valid. As a consequence, the reconstruction of the conductivity profile $\sigma(z)$ can be doubted since many functions $\sigma$ will lead to the same peak structure. \\

\section{Case study: Si(111)$\sqrt{3}\times\sqrt{3}$-Ag\label{caseStudy}}
\begin{figure*}
\includegraphics[scale=1]{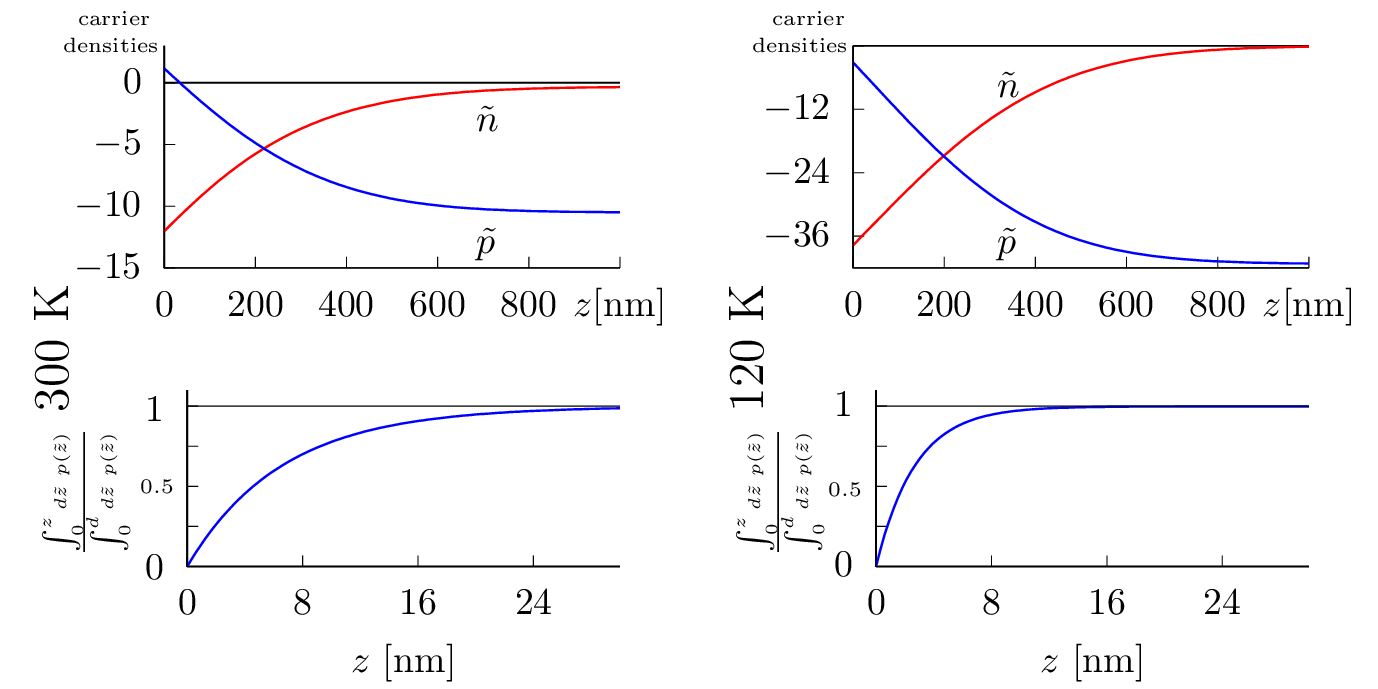}
\caption{Upper panel: Logarithmic variation of the equilibrium electron and hole densities normalized by the bulk carrier density, $\tilde{n}=\log_{10}\tfrac{n_{0}(z)}{p_{b}+ n_{b}}$ and $\tilde{p}=\log_{10}\tfrac{p_{0}(z)}{p_{b}+ n_{b}}$ at 300~K and 120~K. The bulk carrier density reads $6\cdot 10^{15}$ and $5\cdot 10^{15}$~cm$^{-3}$ at 300 and 120~K, respectively. Lower panel: The number of holes located between the surface and the depth $z$ as a fraction of all the positive current carriers in the sample. The exact value of the sample thickness $d$ in the range 100--1000 $\mu$m has no impact on the graphs. These graphs demonstrate that half of the holes are located in the 5~nm zone beneath the surface, while nearly all positive charge carriers are located in the 20~nm subsurface zone at room temperature. At 120~K the holes are concentrated even closer to the surface. \label{carriers}}
\end{figure*}

To demonstrate how the theory works, we comment on the current flow measurements on Si(111)$\sqrt{3}\times\sqrt{3}$-Ag as reported in Ref.~\onlinecite{2d2}. At room temperature, both p-type and n-type doped samples were investigated. For p-type doped samples, three-dimensional currents were observed, while for n-type doped samples, two-dimensional currents were seen. This was explained as a result of the surface Fermi level pinning 0.16 eV above the top of the valence band and the formation of inversion and accumulation layers for n-type and p-type doped samples, respectively. As such, C.~Liu \emph{et al.} assumed that the two-dimensional current was flowing through the surface channel. The system has been experimentally investigated in many ways in the context of the surface conductivity~\cite{popular2,costam}. There is no doubt that the current through the surface states depends also on the bulk electronic structure, as evidenced by the fact that the two-dimensional conductance coincides with the appearance of the inversion layer. As a consequence, if the surface band is an active transport channel, it is not known what fraction of the current goes through. We have no good model of the phenomenon, so we assume a limiting case---the inversion layer as the dominant channel. Reports on Ge(001) and Ge(001):H~\cite{JAP} demonstrate that it is always a viable option. \\
The bulk Fermi level in the sample is said~\cite{2d2} to be about 0.25--0.30~eV below the conduction band at room temperature. This is consistent with a low donor density, in calculations we assume $6\cdot 10^{15}$~cm$^{-3}$. Four point probes were arranged in a line with an equal spacing of 20~$\mu$m. In that geometry, the change to the resistance in the temperature range 120--300~K was measured. The resistance decreased along with the temperature from 300~K to about 140~K; however, at lower temperatures it remained roughly constant. The measurements were performed on samples of 500 $\mu$m in thickness.\\
We begin by calculating the equilibrium quantities: $V_{0}$, $n_{0}$, and $p_{0}$; see sect.~\ref{model}. 
The equilibrium densities of the carriers at 120 and 300~K are shown in Fig.~\ref{carriers}. The hole density at the surface highly exceeds the electron one, so we neglect the electron current. Notably, the carrier density at the surface is not governed by the Debye--H{\"u}ckel length, which shows how the electrostatic potential changes. \\
Next, we calculate the function $\psi^{2}(k;0)$. The peak for $k\sim 0$ appears to be extremely thin. Due to the accumulation of numerical errors, we cannot accurately analyze its structure. The data for $k>10^{-6} \ \mu$m$^{-1}$ suggest a single peak with $\Gamma \sim 10^{-10} \ \mu$m$^{-1}$; more sophisticated numerical work could probably alter this value but cannot make $\Gamma^{-1}$ commensurate with the sample thickness $d$. This makes it necessary to reanalyze the model in terms of discrete eigenvalues and eigenfunctions; see Appendix~\ref{finite}. The result can be easily stated. What counts is the zero mode only. The contribution of the mode associated with the lowest non-zero eigenvalue $k_{1}^{2}$ is proportional to $\psi^{2}(k_{1},0)$. Since $k_{1}^{2}\sim d^{-2}$, $\psi^{2}(k_{1};0)$ is orders of magnitude smaller than the zero mode. The electrostatic potential reads as
\begin{equation}\label{zeroMode}
v(r,z=0)=\frac{I v_{2}(0)}{2\pi \int_{0}^{d}dz\ \sigma(z)}\ln\left(q r\right), 
\end{equation}
where $q$ is a dummy parameter, as discussed in sect.~\ref{reduction}. The zero mode usually prevails at a distance from the source comparable to the sample thickness. Here, it is the conductivity profile that promotes that mode to the only measurable mode. Note that it explains why two-dimensional currents are detected at distances from several micrometers to half a millimeter~\cite{2d2} with no substantial change to the resistivity.\\
Following eq.~(\ref{lincur2}) the current density $\mathbf{j}_{p}$ reads $\mu_{p}p_{0}\nabla\phi$. Under the assumption that the mobility is a single number, i.e. it does not vary with $z$, the current density in the radial direction is proportional to $p_{0}(z)/r$ and vanishes in the other directions. So, the largest current density is close to the surface and it rapidly decreases with the distance from the surface; see Fig.~\ref{carriers}. In our approach, the conductivity of the inversion layer is decoupled from the bulk conductivity. It is in stark contrast to the models using the drift equation that assume a continuous transformation of the conductivity between the hole conductivity at the surface and the electron one in the bulk~\cite{nowe}. \\
\begingroup
\squeezetable
\begin{table}[b!]
\begin{tabular}{l||c|c|c|c|c|c|c|c|c|c} 
T [K] & 120 & 140 & 160 & 180 & 200 &220 &240 &260& 280& 300 \\
\hline
$\mu_{p}$ & 242 & 201& 151&97.7 &61.5 &34.5 &17.2 &8.33 &4.07 &2.68\\
\hline
$\mu_{p}^{v=1}$ & $9\cdot 10^{5}$& 66 252& 7 887&1 235 &277 &79.0 &27.0 &10.6 &4.64 &2.95
\end{tabular}
\caption{Hole mobility calculated from experimental data $\mu_{p}$ [cm$^{2}$/(V$\cdot$s)] using the drift-diffusion scheme and $\mu_{p}^{v=1}$ [cm$^{2}$/(V$\cdot$s)] from the drift equation, i.e., under the assumption $v_{2}(0)=1$. \label{mobil}}
\end{table}
\endgroup
The use of the drift-diffusion equations also has quantitative consequences. This is evident for the mobility values that we can calculate from the resistance $R$ measured in the experiment~\cite{2d2} and our model. The relevant formula reads as
\begin{equation}
 R=\frac{\ln 2}{2\pi\mu_{p}}\frac{v_{2}(0)}{\int_{0}^{d}dz \ p_{0}(z)},
\end{equation}
where $v_{2}$ is defined by eq.~(\ref{potonedim}). The results are shown in Tab.~\ref{mobil}. The obtained mobility at room temperature is extremely small, but surprisingly, the values on the order of 10 cm$^{2}$/(V$\cdot$s) were reported in Ref.~\onlinecite{mobili} for the surface channel on the Si(111)-$\sqrt{3}\times\sqrt{3}$-Ag. The variation of the mobility with temperature is reasonable within the drift-diffusion equations; the change calculated from the drift equation seems to be nonphysical. The mobility in inversion layers has been investigated for a long time~\cite{hansch}, mostly in the context of microelectronics~\cite{ieee}. A further study is necessary to determine if the mobility calculated here can be simply related to those results. \\
\section{Conclusions}
In this paper, we described the surface-sensitive transport measurements by the drift-diffusion equations. The outlined calculation scheme offers a unified and logical framework. The space charge layer corresponds to the equilibrium solution, while the transport equations are regarded as a small perturbation of those solutions. We qualitatively characterize the screened Poisson equation needed to calculate the electrostatic potential. This allows us to draw a connection between our model and the models using the drift equation. We also analyze the mechanism leading to two-dimensional currents. It reveals that the surface currents are governed by very few parameters for the accumulation and inversion layers. It is unclear if the Poisson equation can help in reconstructing the conductivity profiles. Finally, we analyze the experimental data for the Si(111)-$\sqrt{3}\times\sqrt{3}$-Ag within a simple model of the inversion layer. In this case, the observed two-dimensional currents are due to the zero mode of the minority charge carriers.\\
The work can be continued in many ways. To develop a complete model comparable to experiments, the incorporation of the surface bands is imperative. Another interesting issue is the universality of the Lorentzian shape of the function $\psi^{2}(k;0)$. Both the mathematical mechanism resulting in this shape and an explicit formula for $\Gamma$ are highly anticipated. The  limits for the reconstruction of the conductivity profile, which are approximately outlined in our work, need further elaboration. Finally, the implication of the close-to-surface mobility (as calculated above) and its transferability between various systems is an open issue that can be highlighted by future experiments. 

\acknowledgements
The author thanks Professor J. Konior for reading the manuscript, and anonymous referees for insightful discussions and useful references. 
Funding for this research has been provided by the EC under the Large-scale Integrating Project in FET Proactive of the 7th FP entitled ``Planar Atomic and Molecular Scale devices'' (PAMS). 

\appendix
\section{Relevant solutions \label{poissII}}
Based on the developments in a previous paper~\cite{PRB} and in sect.~\ref{reduction}, we know that the drift equation, and hence the current equations for the quasi-Fermi level, can produce two- and three-dimensional currents. Here, we calculate the electrostatic potential for two relevant profiles of the quasi-Fermi levels. First, we consider two-dimensional currents with the quasi-Fermi level given by the formula
\begin{equation}\label{imref2d}
\varphi_{p}(r,z)=\frac{I}{2\pi \sigma_{2}}\ln{q r},
\end{equation}
where $I$ is the current supplied by the source, $\sigma_{2}$ stands for the two-dimensional conductance, and $q$ stands for a (dummy) parameter making the argument of $\ln(\cdot)$ dimensionless. We consider the current flowing in one channel only, to be specific, the hole current. We plug the Ansatz
\begin{equation}\label{ansatz}
v(r,z)=\frac{I}{2\pi\sigma_{2}}v_{2}(z) \ln{qr}
\end{equation}
into the screened Poisson equation~(\ref{poisson}) and obtain the relation for $v_{2}$:
\begin{equation}\label{potonedim}
\left[-\frac{d^{2}}{dz^{2}}+\lambda\left(n_{0}(z)+p_{0}(z)\right)\right]v_{2}(z)=-\lambda p_{0}(z), 
\end{equation}
The boundary condition at the surface is $\tfrac{d}{dz}v_{2}(0)=0$. The solution can be formulated using the appropriate Green's function. Alternatively, one can solve this equation requiring that $v_{2}(z)\to p_{0}(z)/\left[n_{0}(z)+p_{0}(z)\right]$ for $z\to \infty$. These two methods are equivalent and ensure that the solution does not couple to the unbounded (nonphysical) solutions of eq.~(\ref{potonedim}). Note that the potential is $z$-dependent, contrary to the quasi-Fermi level $\varphi_{p}$. Following eq.~(\ref{ansatz}), the surface electrostatic potential has the form
\begin{equation}
v(r,z=0)=\frac{I}{2\pi\sigma_{2}}v_{2}(0) \ln{q r},
\end{equation}
which we use in sect.~\ref{presentation} and~\ref{caseStudy}. \\
Analogously, for the surface quasi-Fermi level behaving as $r^{-1}$, one can assume that the quasi-Fermi level has the form 
\begin{displaymath}
\varphi(r,z)=\frac{I}{2\pi\sigma}\frac{1}{\sqrt{r^{2}+z^{2}}},
\end{displaymath} 
with $\sigma$ standing for some three-dimensional conductivity. The Ansatz
\begin{displaymath}
v(r,z)=\frac{I}{2\pi\sigma}\frac{v_{3}(z)}{\sqrt{r^{2}+z^{2}}}
\end{displaymath}
allows writing the Poisson equation in the form of eq.~(\ref{potonedim}). However, in this case it is an approximate equation, neglecting terms of the order $z/(r^{2}+z^{2})^{3/2}$, and hence valid close to the surface only. The local nature of the screened Poisson equation suggests that the solution at the surfaces is close to the all-inclusive solution. \\
 
\section{Boundary conditions}\label{app}
In Ref.~\onlinecite{PRB} it was noted that the condition $\tfrac{d}{dz}\sigma(0)\neq 0$ might result in the asymptotic behavior $\phi(r) \sim r^{-3}$ for large values of $r$. It was based on the observation that generalized wavefunctions $\{cos(kz +\Phi(k))\}_{k>0}$, where 
\begin{equation}
\Phi=-sign\left(\frac{d\sigma(0)}{dz}\right)\arccos{\frac{k}{\sqrt{k^{2} + \left(\frac{1}{2\sigma(0)}\frac{d\sigma(0)}{dz}\right)^{2}}}},
\end{equation}
vanish for $z=0$ and $k\to 0$. The long-range behavior of the current is given by $\psi(0;0)$. Indeed, if $\psi(0;0)=0$ there are no $1/r$ terms in the Green's function of operator $\hat{L}$, see eq.~(\ref{integ}). However, it is not the case here. It follows from the presence of the zero-resonance $\sqrt{\sigma(z)\sigma^{-1}(\infty)}$. This function satisfies the boundary condition~(\ref{boundaryTransformed}) and solves eq.~(\ref{1dim}) with $ k=0$; factor $\sigma^{-1}(\infty)$ ensures the correct normalization. As such, it locally describes how the solutions satisfying the same boundary condition as $\sqrt{\sigma(z)\sigma^{-1}(\infty)}$ behave near $z=0$ for small $k$. Hence, $\psi^{2}(0;0)= \sigma(0)\sigma^{-1}(\infty)$ and the usual asymptotic behavior $\varphi(r)\sim(\sigma(\infty)r)^{-1}$ holds for $r\to \infty$. The arguments given in Ref.~\onlinecite{PRB} for $k\to\infty$ remain valid in the case at hand as $\cos^{2}\Phi(k)\to 1$ for $k\to\infty$.\\

\section{Absence of bound eigenstates of $\hat{L}$\label{opL}}
In this section, we argue that the current operator $\hat{L}$, see eq.~(\ref{operatorL}), admits no negative eigenvalues. Solutions of eq.~(\ref{start}) with no source correspond to extrema of the functional 
\begin{equation}\label{en1}
E[\varphi]=\frac{1}{2} \int d\Omega \ \sigma \left(\nabla\varphi\right)^{2},
\end{equation}
which is interpreted as the energy dissipated by the current in the source-free region of space per unit time~\cite{Landau} when $\varphi$ stands for the voltage drop (within the drift equation).
The integration is done over the whole region. For any given function $\varphi$, the functional results in a non-negative number. Upon substitution~(\ref{transform}) the energy gains an equivalent form
\begin{equation}\label{en2}
E[\xi]=\frac{1}{2} \int d\Omega \ \xi\hat{L}\xi,
\end{equation}
excluding any negative eigenvalue as well as the zero eigenvalue with a square-integrable eigenfunction. The equivalence between functional~(\ref{en1}) and~(\ref{en2}) can be shown if boundary condition~(\ref{boundaryTransformed}) is taken into account. In the cases of interest here, there are two options. For the semi-space approximation $\mathbb{R}^{2}\times[0,\infty)$, the potential $U(z)$ defined by eq.~(\ref{potential}) asymptotically approaches zero, $U(z)\to 0$ for $z\to\infty$. Hence, no bound states are allowed and only a continuous (scattering) spectrum is present. Thus, the conclusions of Ref.~\onlinecite{PRB} are valid in any case; the current will have the three-dimensional character at sufficiently small and large distances from the source. Therefore, the confinement close to the surface may only appear at limited distances. The case of a real sample of a finite thickness ($\mathbb{R}^{2}\times[0,d]$) is described in Appendix~\ref{finite}.\\

 \section{Depth independence of the quasi-Fermi level\label{const}}
Here, we aim to show that the logarithmic behavior of the quasi-Fermi level is accompanied by the depth-independence of the quasi-Fermi level, and rationalize the Ansatz~(\ref{ansatz}). 
Consider solutions of the equation 
\begin{equation}
\hat{L}u(k;z)=k^{2}u(k;z)
\end{equation}
with the boundary conditions 
\begin{eqnarray}
 u(0)=\sqrt{\frac{\sigma(0)}{\sigma(\infty)}}, \\
 \left.\frac{d u(z)}{dz}\right|_{z=0}=\left. \frac{d }{dz}\sqrt{\frac{\sigma(z)}{\sigma(\infty)}}\right|_{z=0}. 
\end{eqnarray}
For small $k$, functions $u$ approach pointwise the solution $\sqrt{\tfrac{\sigma(z)}{\sigma(\infty)}}$. Hence, for $0\leq z\leq z_{0}$ one can approximate 
\begin{equation}\label{approxC}
u(k;z)\approx \sqrt{\tfrac{\sigma(z)}{\sigma(\infty)}}.  
\end{equation}
 Numerical investigations show that for a class of physical $\sigma(z)$ this approximation is valid for $z_{0}$ in the range of the surface layer. The difference between $u(k;z)$ and the (generalized) eigenfunctions $\psi(k;z)$ lies in normalization. The multiplicative factor can be easily found
\begin{equation}\label{approximD}
u(k;z)=\sqrt{\frac{\sigma(0)}{\sigma(\infty)}}\frac{\psi(k;z)}{\psi(k;0)}.
\end{equation}
Combining the approximations~(\ref{approxC}),~(\ref{approximD}), and the Green's function formula~(\ref{green}) we obtain 
\begin{equation}
\varphi(r,z)=\frac{I}{2 \pi} \int dk \ K_{0}(kr) \psi^{2}(k;0),
\end{equation}
which does not depend on $z$ and motivates the Ansatz~(\ref{ansatz}).\\
 
 \section{Finite-thickness effects\label{finite}}
For the current equation~(\ref{start}) considered for samples of finite thickness $d$, the expansion of the Green's function in terms of the eigenfunctions and eigenvalues reads as
\begin{equation}
G(\mathbf{x};\mathbf{x'})=\frac{1}{2 \pi \sqrt{\sigma(z)\sigma(z')}}\sum_{n=1}^{\infty} \ K_{0}(k_{n}r) \psi(k_{n};z)\psi(k_{n};z'),
\end{equation}
where $k_{n}^{2}$ stands for the $n$-th eigenvalue of the operator $\hat{L}$; see eq.~(\ref{operatorL}). In Sect.~\ref{caseStudy}, we consider the system that strongly enhances the modes close to $k=0$. The width $\Gamma$ of the peak in $\psi^{2}(k;0)$ is orders of magnitude smaller than $k_{1}\sim d^{-1}$. Hence, the above modes cannot describe the surface currents. However, to any solution of eq.~(\ref{start})  zero modes can be added. Far from the source, the zero mode has the form
\begin{equation}
C+\frac{1}{\int_{0}^{d}d\tilde{z} \ \sigma(\tilde{z})} \ln{q r},
\end{equation}
where $C$ is an arbitrary constant and $q$ is the dummy parameter. Note that $q$ and $C$ are not independent. Heuristically, the logarithm in the above formula is due to summing up all modes parallel to the surface with the zero eigenvalue in the normal direction, i.e.,
\begin{equation}
\begin{split}
\frac{\ln q\sqrt{(x-x')^{2}+(y-y')^{2}}}{\int_{0}^{d}d\tilde{z} \ \sigma(\tilde{z})}-C = \frac{\psi(0;z)\psi(0;z')}{2 \pi \sqrt{\sigma(z)\sigma(z')}} \times \\ \times \int_{-\infty}^{\infty}dk_{x} \int_{-\infty}^{\infty}dk_{y} \ \frac{\exp{i\left(k_{x} (x-x')+k_{y}(y-y')\right)}}{k_{x}^{2}+k_{y}^{2}}.
\end{split}
\end{equation}
As such, we expect that only the zero mode $\sim \ln{q r}+C$ contributes to the measured current. This approach supports the conclusions of Appendix~\ref{const} that quasi-Fermi levels associated with the logartithmic currents are $z$-independent. Furthermore, it allows for a non-rigorous estimation of parameter $\Gamma$ by comparing the above coefficient of $\ln(\cdot)$ function and eq.~ (\ref{sigma2}). It results in the relation
\begin{equation}
\Gamma\sim \frac{\sigma(\infty)}{\int_{0}^{d}d\tilde{z} \ \sigma(\tilde{z})},
\end{equation}
which may be used if the resulting $\Gamma$ does not depend on $d$ over several orders of magnitude.

\end{document}